\RequirePackage{fix-cm}
\documentclass[smallextended]{svjour3}       
\smartqed  
\usepackage[ngerman, english]{babel}
\usepackage{amssymb,amsfonts,amsmath,amscd,bbm,mathrsfs}
\usepackage{xcolor}
\usepackage{pdfcolmk}
\usepackage{pgf,pgfplots}
\usepackage{graphicx,wrapfig}
\usepackage{hyperref}
\usepackage{mathdots,scalerel,subcaption, multirow}
\usepackage{float}
\usepackage{natbib}
\usepackage{amsopn}

\newcommand{\E}{{\mathbb E}}

\newcommand{\N}{{\mathbb N}}
\newcommand{\Q}{{\mathbb Q}}
\newcommand{\R}{{\mathbb R}}

\newcommand{\sref}[2]{\hyperref[#2]{#1~\ref*{#2}}}
\newtheorem{rem}{Remark}

\makeatletter

\usepackage{url}

\usepackage{float}
\usepackage{tikz}

\definecolor{Darkgreen}{rgb}{0,0.7,0}
    \newcommand{\darkgreen}{\color{Darkgreen}}
\usetikzlibrary{patterns}

\journalname{submitted to journal}
\begin{document}

\title{A deep learning model for gas storage optimization
}
\subtitle{}


\author{Nicolas~Curin \and
        Michael~Kettler \and
        Xi~Kleisinger-Yu \and
        Vlatka~Komaric \and
        Thomas~Krabichler \and
        Josef~Teichmann \and
        Hanna~Wutte \and
}


\institute{N.~Curin \at Axpo~Solutions~AG$^\ast$, Baden, \email{nicolas.curin@axpo.com} \and
           M.~Kettler \at Axpo~Solutions~AG$^\ast$, Baden, \email{michael.kettler@axpo.com} \and
           X.~Kleisinger-Yu  \at Department of Mathematics, ETH~Z\"urich, \email{xi.kleisinger-yu@math.ethz.ch} \and
           V.~Komaric  \at Axpo~Solutions~AG$^\ast$, Baden, \email{vlatka.komaric@axpo.com} \and
           T.~Krabichler \at Eastern Switzerland University of Applied Sciences, St.~Gallen,
           \email{thomas.krabichler@ost.ch} \and
           J.~Teichmann \at Department of Mathematics, ETH~Z\"urich, \email{josef.teichmann@math.ethz.ch} \and
           H.~Wutte \at Department of Mathematics, ETH~Z\"urich, \email{hanna.wutte@math.ethz.ch} \and
           $^\ast$ Opinions expressed in this paper are those of the authors, and do not necessarily reflect the view of Axpo~Solutions~AG.
}

\date{Received: date / Accepted: date}

\maketitle

\begin{abstract}
To the best of our knowledge, the application of deep learning in the field of quantitative risk management is still a relatively recent phenomenon. In this article, we utilize techniques inspired by reinforcement learning in order to optimize the operation plans of underground natural gas storage facilities. We provide a theoretical framework and assess the performance of the proposed method numerically in comparison to a state-of-the-art least-squares Monte-Carlo approach. Due to the inherent intricacy originating from the high-dimensional forward market as well as the numerous constraints and frictions, the optimization exercise can hardly be tackled by means of traditional techniques.
\end{abstract}

\section{Introduction}
Natural gas prices exhibit distinct yearly seasonal patterns. Due to limited storage capacities and pronounced fluctuations in the demand, its prices tend to be lower in summer, and higher as well as more volatile in winter. Physical storage facilities are required in order to exploit this seasonality. Various market participants are willing to own or lease storage facilities, or trade storage capacities, which creates high demand in the underground gas storage facilities worldwide. As an example, US working gas\footnote{Working gas refers to the total volume of gas in storage at a particular point in time. It is computed as total gas volume minus base gas.} in underground storage was at its record high in 2020 compared to the previous five years; see \sref{Figure}{Fig_EIA}. Given the smoothing effect of gas price spreads through storage facilities, it is vital to optimize their usage for trading, pricing, hedging, risk management and investment purposes.

\begin{figure}[h]
\begin{center}
\includegraphics[page=4,trim={11.04cm 4.5cm 3.3cm 21.7cm},clip,width=0.5\textwidth]{pic_20210218_natural_gas_storage_dashboard}
\includegraphics[page=1,trim={1.4cm 12.1cm 11.13cm 12.4cm},clip,width=0.5\textwidth]{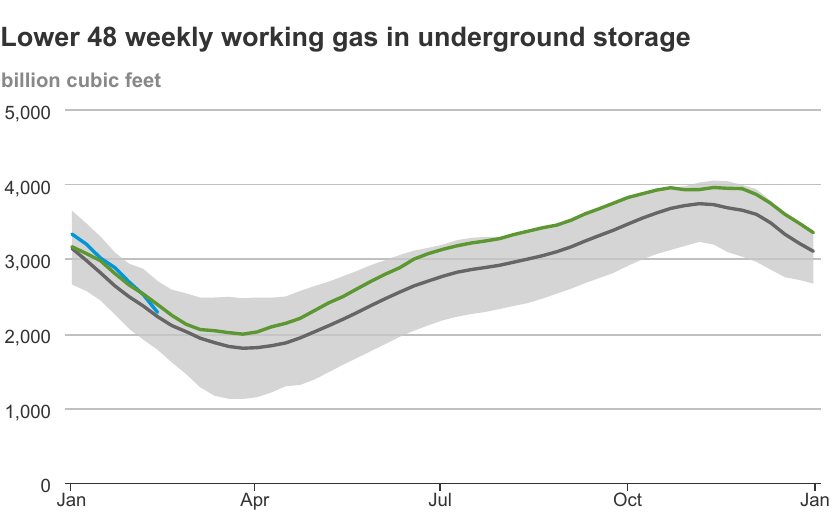}
\caption{US natural gas (in contrast to liquefied natural gas or briefly LNG) futures curve and storage information provided by the EIA (US Energy Information Administration), as per February~18,~2021; downloaded from \url{https://www.eia.gov/naturalgas/storage/dashboard/}. The upper plot shows one year natural gas futures curves consisting of twelve monthly futures contracts with delivery period of months ranging from March~2021 to February~2022. It features the above-mentioned seasonal pattern, namely higher prices in winter and lower prices in summer. The lower plot shows the lower bound of underground gas storage in the US in lower 48 weekly working gas. Gas storage in the year 2020 reached a record high compared with that of the previous five years.}\label{Fig_EIA}
\end{center}
\end{figure}

Over the last decades, various articles contributed to the modeling and optimization of energy storage. For standard references, see the Section~12.6 of \cite{geman2009commodities}, 
Section~5.3.4 of \cite{fiorenzani2012handbook}, and \cite{holland2007optimization, holland2008decision}. Other references include, e.g., \cite{de2015gas, boogert2008gas,safarov2017natural,cummins2017gas, carmona2010valuation,bjerksund2011gas, thompson2009natural,henaff2018gas,malyscheff2017natural}. Much of the literature puts more emphasis on the modeling (and prediction) of gas prices rather than on developing algorithms for the optimization of storage plans. Conventionally, storage plans were optimized by means of Least-Squares Monte~Carlo approaches (LSMC) \citep{malyscheff2017natural} or support vector machine regression (SVR), considered as a stochastic control problem with HJB equations \citep{THOMPSON201626}, or an application of real option theory \citep{thompson2009natural}. The bottleneck of classical optimization techniques is the so-called \emph{curse of dimensionality}, i.e., the running time often grows exponentially in the number of state space dimensions. Based on techniques \emph{inspired by} reinforcement learning, one manages to tackle these intricate optimization tasks without simplifications. To this end, one designs an artificial financial agent with superhuman experience and a decent risk appetite, who is able to trade off numerous aspects without further ado. The recent work \citet{bachouch2020deep} applies a number of true reinforcement learning algorithms to the problem of gas storage valuation, seen as discrete-time stochastic control problems in finite time horizon. Regarding the problem of pricing commodity swing options, \cite{daluiso2020pricing} employ an actor-critic reinforcement learning technique to approximate actions in day-ahead forward markets maximizing accumulated expected payoffs.
In contrast to these reinforcement learning methods, our artificial agent does not learn
to act under \emph{all} configurations, but only under those relevant for the given scenarios. Furthermore neither Markov assumptions are made, nor is dynamic programming used.

In spirit similar to the present work, \citet{barreraesteve:inria-00117175} suggest to address the problem of pricing a swing option on natural gas with a policy search method, i.e., to train a neural network to approximate optimal gas consumption rates maximizing expected terminal wealth. Thus, in accordance with our approach, \citet{barreraesteve:inria-00117175} view the task of pricing the swing option as a general parametric optimization problem rather than one of stochastic control. By contrast however, we do not assume any Markov setting such as the one arising from the one-factor model for forward prices of gas that is considered therein. Moreover, we stress that the deep hedging approach followed in the present work can be readily generalized to minimizing risk measures instead of expected rewards.

Beyond that, to the best of the authors' knowledge, our techniques inspired by reinforcement learning have not been applied to gas storage and related problems. Thus, their full potential as well as new challenges in storage-related optimization problems are yet to be investigated.

In this article, we present a fresh machine learning approach for the optimization of gas storage. Along the lines of deep hedging (see \cite{buehler2019deep}), we determine optimal \emph{strategy networks}, i.e., neural networks that approximate optimal strategies for trading in spot and forward markets utilizing storage facilities. Optimality is understood as maximizing expected utility of accumulated wealth at terminal time. More specifically, we introduce two models which are of the intrinsic valuation type: a simple \emph{spot-only model} (SMod) allowing for trades in a spot-proxy only, and a more involved model referred to as \emph{spot-and-forward model} (SFMod) additionally incorporating trades in monthly forwards with delivery periods. Traditionally, models with trades in spot-proxies based on an artificial daily forward curve, which is implied from the tradable monthly forwards with delivery period, have been employed for simplicity. However, the main purpose of gas storage optimization is to maximize profits or utility of storage managers rather than theoretical valuations. Therefore, the use of tradables, and thus the use of SFMod, is more relevant for gas storage optimization.

The paper is structured in the following way: In Section~2, we provide a brief overview of important aspects in gas storage modeling and outline the machine learning concept that we employ for optimizing gas storage usage. In Section~3, we present the spot-only deep learning model (SMod) for trading strategies utilizing gas storage facilities. We compare our model in numerical tests against a set of benchmark strategies derived via LSMC. In Section~4, we present the spot-and-forward model (SFMod), which additionally includes trades in monthly forwards with delivery periods, and investigate its performance in numerical tests likewise.

Throughout this article, we consider a discrete time setting. Let the time instances $\mathbb{T}=\{0,1,2,\hdots,K-1\}$ for some $K\in\mathbb{N}$ be the trading horizon in days, and let $(\Omega,\mathcal{F},\mathbb{F},\mathbb{P})$ with $\mathbb{F}=(\mathcal{F}_k)_{k\in\mathbb{T}}$ be a filtered probability space with real-world measure $\mathbb{P}$. Additionally, we assume the existence of an equivalent risk-neutral measure $\Q$.

\section{Optimizing gas storage by means of deep hedging}\label{sec:optimizinggasstorage}

There are three types of underground natural gas storage: depleted natural gas field/oil fields, aquifers, and salt caverns. They are often located close to a pipeline, which makes the delivery of physical transactions more convenient. Compared to the storage through conversion to LNG, underground natural gas storage is bigger and cheaper, but restricted to regional use. In \sref{Table}{Tab_storage}, we provide an overview of the main stylized characteristics of a gas storage, that are relevant for modeling and optimization.  
\begin{table}[h]
\begin{center}
\resizebox{0.6\columnwidth}{!}{%
\begin{tabular}{c|c}
\multicolumn{2}{c}{ storage optimization constraints with unit: therm or MWh }\\ \hline \hline
initial storage & 0 units (plus cushion gas) \\ \hline
terminal storage & 0 units (plus cushion gas) \\ \hline
capacity & c \\ \hline
injection rate on day $k$ & $u_k$ units ($u_k >0$) \\ \hline
withdrawal rate on day $k$ & $\ell_k$ units ($\ell_k < 0$) \\ \hline
injection cost & $\kappa\in [0,1]$ \\ \hline
withdrawal cost & $\kappa\in [0,1]$ \\ \hline
overhead (one time expense) & $C \, \$ $ \\ \hline
\end{tabular}}\caption{This table lists the most important characteristics for modeling gas storage. Note that in underground storage, there usually is cushion or base gas, which is the volume of natural gas that is intended as permanent and not withdrawable inventory to maintain minimal pressure. For simplicity, the injection and withdrawal costs are assumed to be proportional to injection and withdrawal respectively. In reality, these costs depend additionally on the pressure in the underground storage, which in turn depends on the level of working gas. When trading storage capacities, the parties often agree to neglect the physical complications.}\label{Tab_storage}
\end{center}
\end{table}

The goal of gas storage optimization is to find an optimal plan for withdrawing and injecting gas into the storage over a certain period of time subject to the above-mentioned constraints. Extracting respectively feeding gas into the storage corresponds to going short or long in the spot market with respect to a certain storage level. Hence, gas storage optimization can be seen as the problem of identifying optimal actions in an uncertain and restricted market environment to maximize an expected terminal utility of accrued wealth. More formally, let us consider an agent trading in a market and let $h_k$ denote the action she takes on day $k$. A trading strategy over the whole trading horizon is then collected in $H=\{h_0,...,h_{K-1}\}$. At maturity, the agent gains utility $U(W_H)$ based on the stochastic terminal wealth $W_H$ that she accrued by trading according to the strategy $H$. The agent seeks to identify an optimal strategy $H^*$ satisfying
\begin{align}\label{eq:RLgoal}
    H^*=\max_H\E_\mathbb{P} [U(W_H)].
\end{align}

Reinforcement learning \citep{sutton2018reinforcement} is a broad and very active area of research, suggesting a plethora of algorithms to solve intricate optimization problems. A very popular strand of deep reinforcement learning focuses on approximating optimal actions by (e.g., feed-forward) neural networks; see Definition \ref{def:NN} below. Neural networks are very suitable for such tasks because of their universality and their efficient trainability. Parameters $\theta$ of these network strategies $G^\theta=\{g^\theta_0,...,g^\theta_{K-1}\}$ are trained to maximize an estimate of the expected terminal utility, i.e., to solve 
\begin{align}
        \max_\theta\E_\mathbb{P} [U(W_{G^\theta})].
\end{align}

\begin{definition}[Feedforward Neural Network]\label{def:NN} Let $L, d_0, d_L\in\N$.
		A feed-forward neural network ${g^\theta:\,\R^{d_0}\to \R^{d_L}}$ is defined as
		\begin{equation}\label{eq:RSN}
		g^\theta(x)=A^{L}\circ\varphi\circ A^{L-1}\circ\ldots\circ\varphi\circ A^{1}(x),\quad
		\end{equation}
		where
		\begin{itemize}
		\item $L\in\mathbb{N}$ is the number of layers ($L-1$ hidden layers),
		\item\label{itm:as:ReLU} $\varphi(\cdot)$ denotes a non-linear activation function that is applied component-wise, e.g., the sigmoid activation function $\varphi(\cdot)=(1+e^{-\,\cdot})^{-1}$, and
		\item $A^{l}, l=1,\ldots,L$ denote affine linear maps in the respective dimensions, whose parameters are stored in $\theta\in\R^q$ for some $q\in\mathbb{N}$.
		\end{itemize}
\end{definition}

Note that in reinforcement learning, it is commonly assumed that environments can be described by (known or unknown) Markov decision processes. However, in many real-world applications and in particular in financial markets, information above and beyond knowledge of the current state can be used to better predict the dynamics of the environment and improve control, rendering Markov assumptions unrealistic.
While enlarging the state space can partially act as remedy, we highlight that the approach outlined above does not necessarily require a Markovian framework. Instead, we approximate trading strategies along the lines of deep hedging without restricting the state space or even restricting ourselves to specific market dynamics.

Finally, for obvious reasons, we want to note that the trading action $h_k$ should only depend on information which is available in the market up until time $k$. This entails the parameterization of the strategy $h_k$ on a given day $k$ by a neural network $g^\theta_k$ mapping appropriate  market information and storage levels to trading actions.

\section{SMod: intrinsic spot trading}\label{ML_ToyI}
Following the machine learning approach outlined in Section \ref{sec:optimizinggasstorage}, we introduce in what follows a deep hedging model for gas storage optimization that is based on trading day-ahead prices of gas. Note that in commodities markets, day-ahead futures or forwards are seen as close proxy of the spot price. Therefore, we tacitly refer trading activities in the day-ahead price of gas to spot trading. For simplicity, we assume no discounting and zero transaction cost, i.e., $\kappa\equiv0$; for more general formulation including costs, see \sref{Remark}{rem_fullstorage}.

Let $S_k=\big(F(k,k+1,k+1)\big)_{k\in\mathbb{T}}$ denote the $\mathbb{F}$-adapted gas spot price, and $h^S_k$ the $\mathcal{F}_k$-measurable action on day $k$. $h^S_k>0$ refers to an injection of $|h^S_k|$ MWh into the storage and $h^S_k<0$ refers to a withdrawal of $|h^S_k|$ MWh from it. A trading strategy over the whole trading horizon is denoted by $\widetilde{H}^S=\{h^S_0,h^S_1,...,h^S_{K-1}\}$ and its terminal value is given by $(\widetilde{H}^S\bullet S)_{K-1}:=\sum_{k=0}^{K-1} h^S_k S_k$. Moreover, the storage level (or working gas) $H_n^S$ on day $n$ is given by $$H_{n}^S:=\sum_{k=0}^{n-1}h^S_k,$$ with an initially empty storage, i.e., $H_0^S:=0$. 

Suppose a storage manager's preferences can be expressed through a (concave, non-decreasing) utility function $U:\mathbb{R}\to\mathbb{R}$. In line with Section \ref{sec:optimizinggasstorage}, she aims to identify strategies maximizing her expected utility of terminal wealth, i.e., to maximize
\begin{align}
\mathbb{E}_\mathbb{P}&\big[U(W_{K-1})\big], \label{Eq_utility}
\intertext{over all eligible $\widetilde{H}^S$, where}
W_{K-1}&:=\sum_{k=0}^{K-1}-h^S_k S_k \label{Eq_Wspot}
\end{align}
denotes the resulting terminal profit and loss\footnote{The negative sign is added in analogy to the direction of the cashflows.} (P\&L). The optimization is subject to the constraints
\begin{align}
H_K^S & = 0, \label{eq_1st} \\
0 \,\leq H_k^S \, \leq\, c, \qquad   &\text{and  }
\qquad\ell_k \, \leq \,h^S_k \, \leq\, u_k,\label{eq:3rd}
\intertext{for all $k\in\mathbb{T}$. The constraint \eqref{eq_1st} says that the storage must be empty at maturity; if one does not adhere to that, a contractually agreed penalty becomes due. Unless gas prices become negative, any profit-seeking agent would comply with \eqref{eq_1st} naturally, since profits can only be generated by disposing of previously stored gas. The daily \eqref{eq:3rd} constraints can be transformed and merged to a single target range. Indeed, one simply imposes}
\widetilde{\ell}_k \, \leq \, h^S_k\, & \leq \,\widetilde{u}_k,\label{Eq_rephrase}
\intertext{where}
\widetilde{\ell}_k:=\max\big\{\ell_k,-H_k^S\big\},\qquad &\text{and}\qquad\widetilde{u}_k:=\min\big\{u_k,c-H_k^S\big\}.\nonumber
\end{align}

Pursuing a strategy learning approach, we approximate each action $h^S_k$ in terms of a deep neural network $g_k$; for ease of readability, we henceforth skip the dependence on the model parameters $\theta$. The inputs to these network strategies are the current spot price $S_k$, the latest storage fill level $H_k^S$ (that iteratively depends on the previous neural networks) and the time $k$, i.e., $g_k = g_k\big(k,H_k^S,S_k\big)$. These $K$ networks are summarized in the storage schedule $\widetilde{G}=\{g_0, ..., g_{K-1}\}$. Note that we allow for parameter sharing amongst the network instances $g_k$, i.e., the number of distinct neural networks $N\in\N$ can be significantly smaller than the number of trading days. In the most extreme case, the present framework allows for modeling each strategy with the same network, i.e., $g_k\equiv g$ for all $k\in\mathbb{T}$. It needs to be noted that the inputs of the neural networks in the numerical tests below were normalized in order to ensure a swift and stable learning process. The model parameters of $\widetilde{G}$ were trained with standard Adam stochastic gradient descent on negative expected utility
\begin{align}
    \mathbb{E}_\mathbb{P}&\left[-U\left(\sum_{k=0}^{K-1}-g_k\big(k,H_k^S,S_k\big) S_k\right)\right],
\end{align}
subject to the constraints mentioned in Section~\ref{sec:optimizinggasstorage}.

\subsection{Training setup}
In the following, we state the precise training setup by the example of exponential utility $ U(\cdot):=(1-e^{-r \cdot})/r $, with risk aversion rate $r\in\R^+$.
\begin{itemize}
\item Training Data: time horizon of storage $\mathbb{T}$, $M$ trajectories of the spot $(S^i_k )_{k\in\mathbb{T};i=1,...,M}$.
\item Training object: storage action (withdrawal or injection rate) $\widetilde{G}$ over the whole storage horizon, that is a neural network consisting of $N\in\N$ ($N\le K$) distinct sub-networks, each of which has $L$ layers. The network's input is time as well as respective spot and storage fill level.
\item Training criterion: minimize an estimate of expected negative utility over batches $B\subset\{1,\ldots,M\}$ of training data, i.e.,
\[\min_{\widetilde{G}(S^i)\in\mathcal{G}^i, \forall i\in B}\frac{1}{|B|} \sum_{i\in B}U\big(W^i_{K-1}\big),\] 
where
\begin{align*}
&W^i_{K-1} :=\sum_{k=0}^{K-1}-g_k(k,H^{S_i}_k,S^i_k) S^i_k,\\
&\mathcal{G}^i=\Big\{\widetilde{G}(S^i) ~\Big| ~H_K^{S^i}=0;\  \widetilde{\ell}_k\leq g_k\big(k,H_k^{S^i},S^i_k\big)\leq\widetilde{u}_k\, ~\text{for}~k\in\mathbb{T}\Big\}.
\end{align*}
\end{itemize}

\begin{rem}\label{rem_fullstorage}
For the full case as described in \sref{Table}{Tab_storage}, where costs $\kappa$ and $C$ are non-zero, terminal profit and loss is given by 
\[W_{K-1}:=\bigg(\sum_{k=0}^{K-1}-h^S_k S_k-\big|h^S_kS_k\big|\cdot\kappa\bigg)-C.\]
The training can be performed analogously.
\end{rem}

\begin{rem}
Of course here a more general path dependence could be considered to deal with possible non-Markovianity.
\end{rem}

For numerical testing, spot curves of gas as well as benchmark strategies were provided by Axpo~Solutions~AG in form of $1\,000\times 351$ matrices, representing $M=1000$ scenarios of $K=351$ trading days. Benchmark strategies had been derived utilising the LSMC technique. We would like to emphasize that the proposed framework is not linked to specific stochastic dynamics of the spot price scenarios, i.e., we can exploit the methodology regardless of the model choice. Therefore, we leave the nature of the scenarios unspecified. If required, one can enrich the base scenarios with arbitrary stress scenarios.

The neural network model was implemented in \textsc{tensorflow.keras} with the sigmoid activation function. The daily constraints $\widetilde{\ell}_k$ and $\widetilde{u}_k$ from \eqref{Eq_rephrase}, and the network-based action $g_k$ were parameterized using the inverse linear transformation from $0$, $1$ and $\frac{g_k-\widetilde{\ell}_k}{(\widetilde{u}_k-\widetilde{\ell}_k)}\in [0,1]$ respectively.  For the final zero storage constraint, we additionally checked for every $k$ that
\begin{align}
    H^S_k \leq \sum_{k+1}^{K-1}\ell_k \label{Eq_final_storage}
\end{align}
prevailed, in order to ensure that an empty final storage $H^S_K=0$ remained reachable. If the condition was violated on any day, the upper action bounds on all subsequent days were overridden by their lower counterparts, forcing a complete withdrawal of storage until maturity. In reality, it is possible to leave a non-empty storage by paying the penalty. Yet, for our modeling, the zero final storage rule was strictly adhered to. \sref{Figure}{fig:constraints} visualizes the constraints: the left plot shows normalized zero storage constraint \eqref{Eq_final_storage}, and the right plot shows daily injection and withdrawal bounds \eqref{Eq_rephrase} of the trained financial agent. The original daily constraints with the two regimes
\[\ell_k=\begin{cases}-600&,\text{ if $k\leq$170},\\-3\,072&,\text{ otherwise},\end{cases}\qquad u_k=\begin{cases}2\,808&,\text{ if $k\leq$200},\\408&,\text{ otherwise}.\end{cases}\]
are still recognizable.

\begin{figure}[ht]
    \centering
    \includegraphics[trim={0cm 0cm 1.5cm 0cm},clip,width=0.46\textwidth]{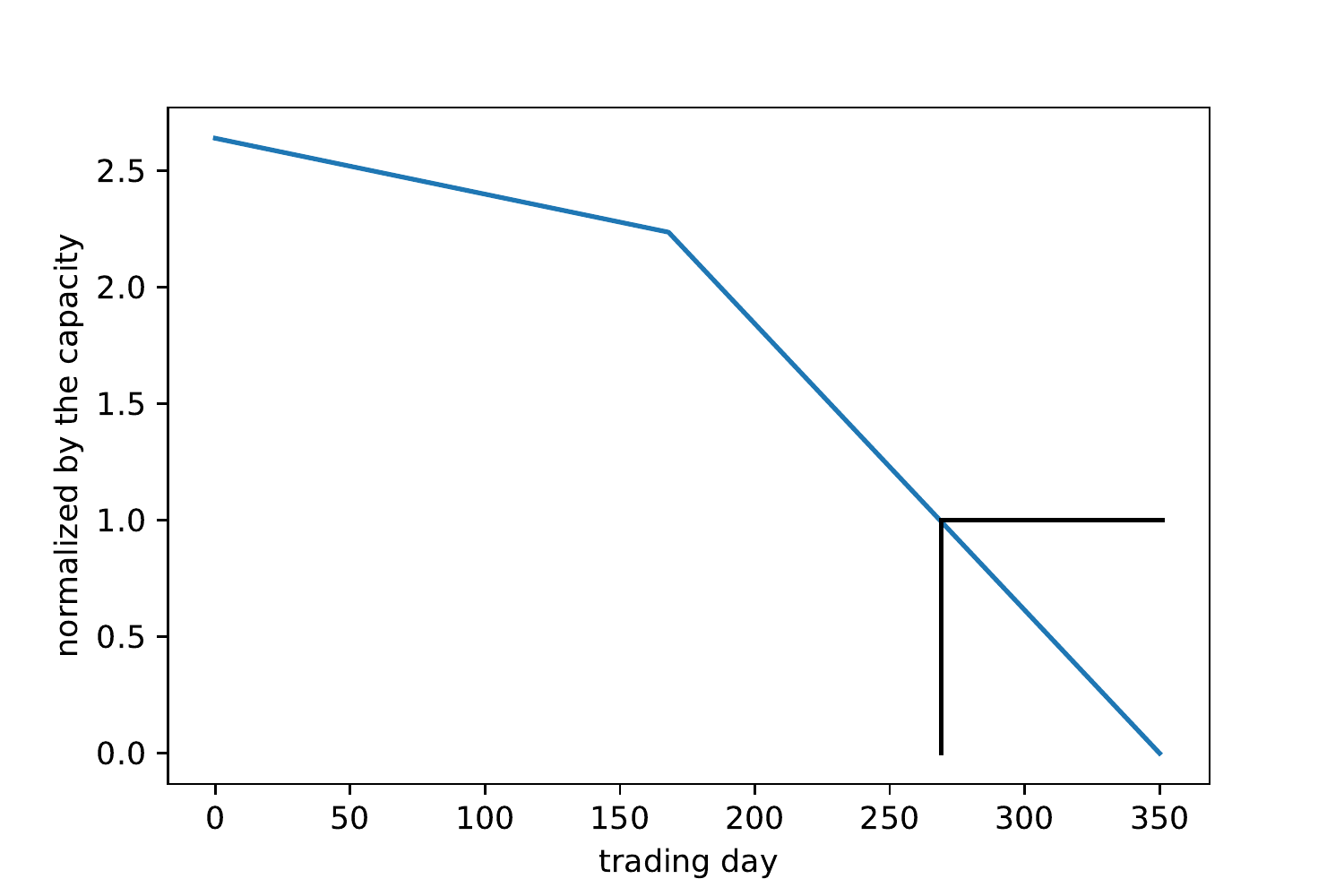}
    \hspace*{0.5em}
    \includegraphics[trim={0cm 0cm 1.5cm 0cm},clip,width=0.46\textwidth]{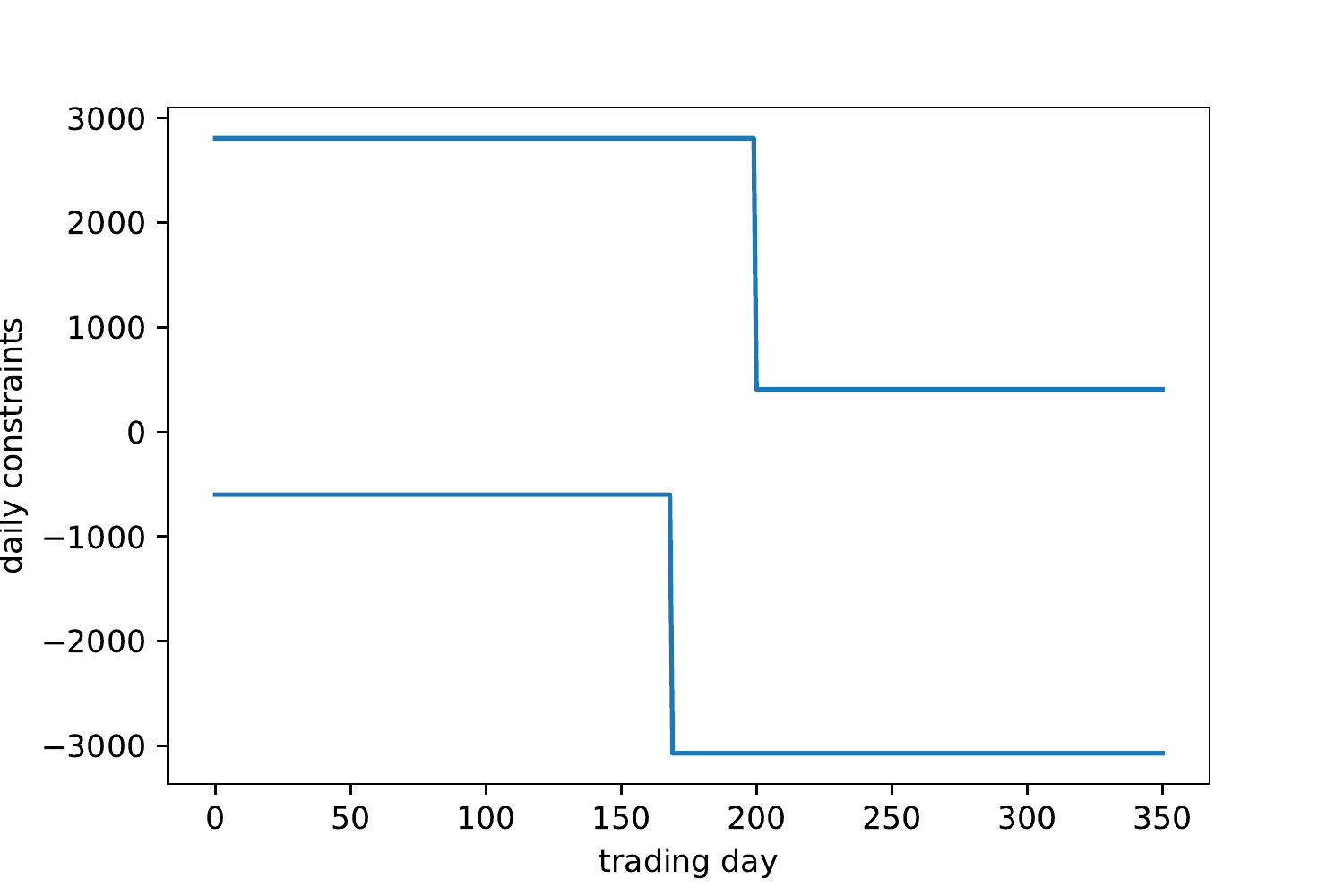}
    \caption{The daily constraints of a gas storage. The left figure visualizes the empty final storage constraint \eqref{Eq_final_storage}; the $y$-axis is normalized by the total storage capacity $c$. The critical boundary is reached on the trading day 269. In other words, if the relative storage level is beyond the blue line any time after the trading day 269, the only admissible actions remain maximal withdrawal up until maturity. The right figure visualizes the daily injection and withdrawal constraints.}
    \label{fig:constraints}
\end{figure}

The network strategies $g_k$ were trained based on the spot prices provided by Axpo~Solutions~AG. The data set was split into a training set of $900$ and a validation set of $100$ scenarios in order to perform in-sample and out-of-sample tests. LSCM benchmark strategies were optimized on the entire set of $1\,000$ scenarios. Thus, they serve as optimal solution. In the course of various experiments, we assessed the required training time depending on the depth and the number of distinct neural networks ($N\leq K$), different learning rates and the batch size in order to fine-tune empirically a suitable setting for SMod. The training generally concluded quickly and is well-managable on a standard 8-core notebook. Illustratively, the training of SMod using in the implementation as much as $K$ neural networks and $1\,000$ epochs on $900$ scenarios takes less than 10~minutes. This runtime is competitive with the LSMC approach. Moreover, it turned out that it is not necessary to build SMod on $K$ neural networks. In fact, we encountered that 12 instances with $L=2$ and $d_1=16$ (see Definition \ref{def:NN} above) already provided a decent approximation of the optimal policy. Indeed, after $1\,000$ epochs on $900$ scenarios, a learning rate of $0.05$, a batch size of $64$, and a risk aversion rate of $r=3$, the strategy of the artificial financial agent gets convincingly close to the benchmark solution. \sref{Figure}{fig:comparison} provides a visualization of the P\&L line-up between the spot-only and the benchmark model in in-sample and out-of-sample tests, as well as a visualization of the storage fill levels of SMod and that of the benchmark respectively. A comparison of descriptive statistics on the terminal P\&L between SMod and the benchmark is reported in the table at the bottom of \sref{Figure}{fig:alphacomparison}.

\begin{figure}[ht]
    \centering
	\includegraphics[trim={0cm 0cm 1.5cm 0cm},clip,width=0.46\textwidth]{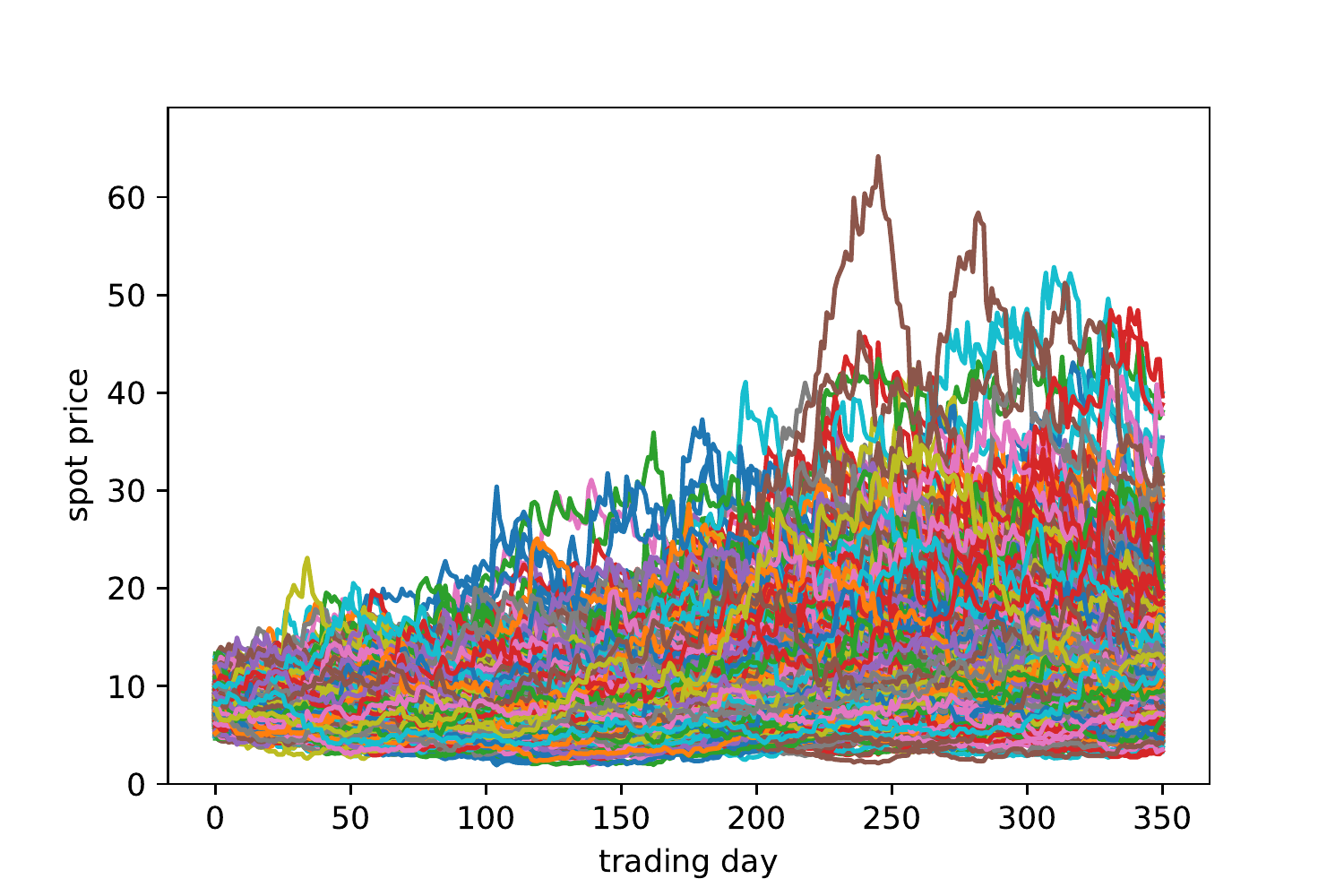}
	\hspace*{0.5em}
  	\includegraphics[trim={0cm 0cm 1.5cm 0cm},clip,width=0.46\textwidth]{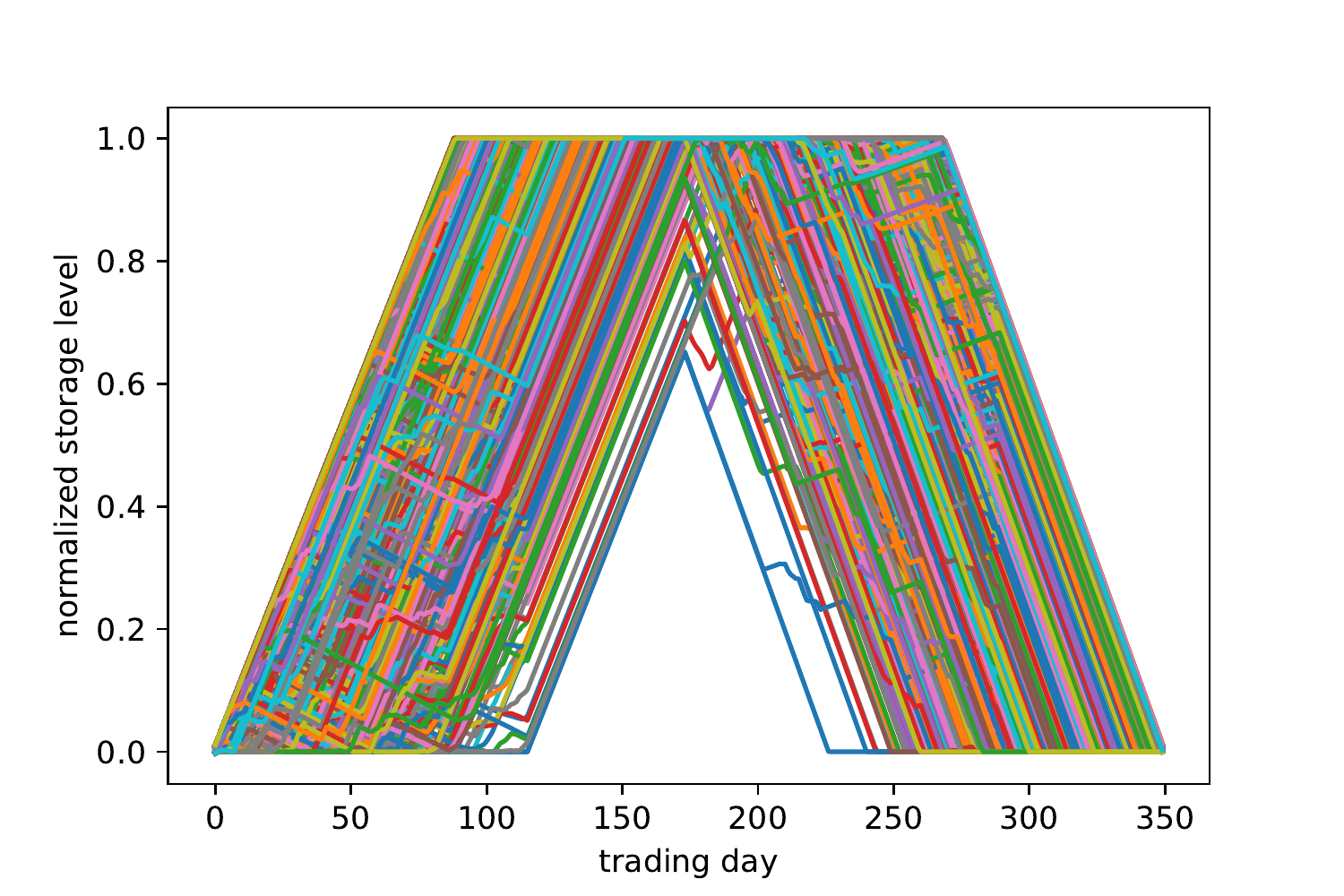}
	\includegraphics[trim={0cm 0cm 1.5cm 0cm},clip,width=0.46\textwidth]{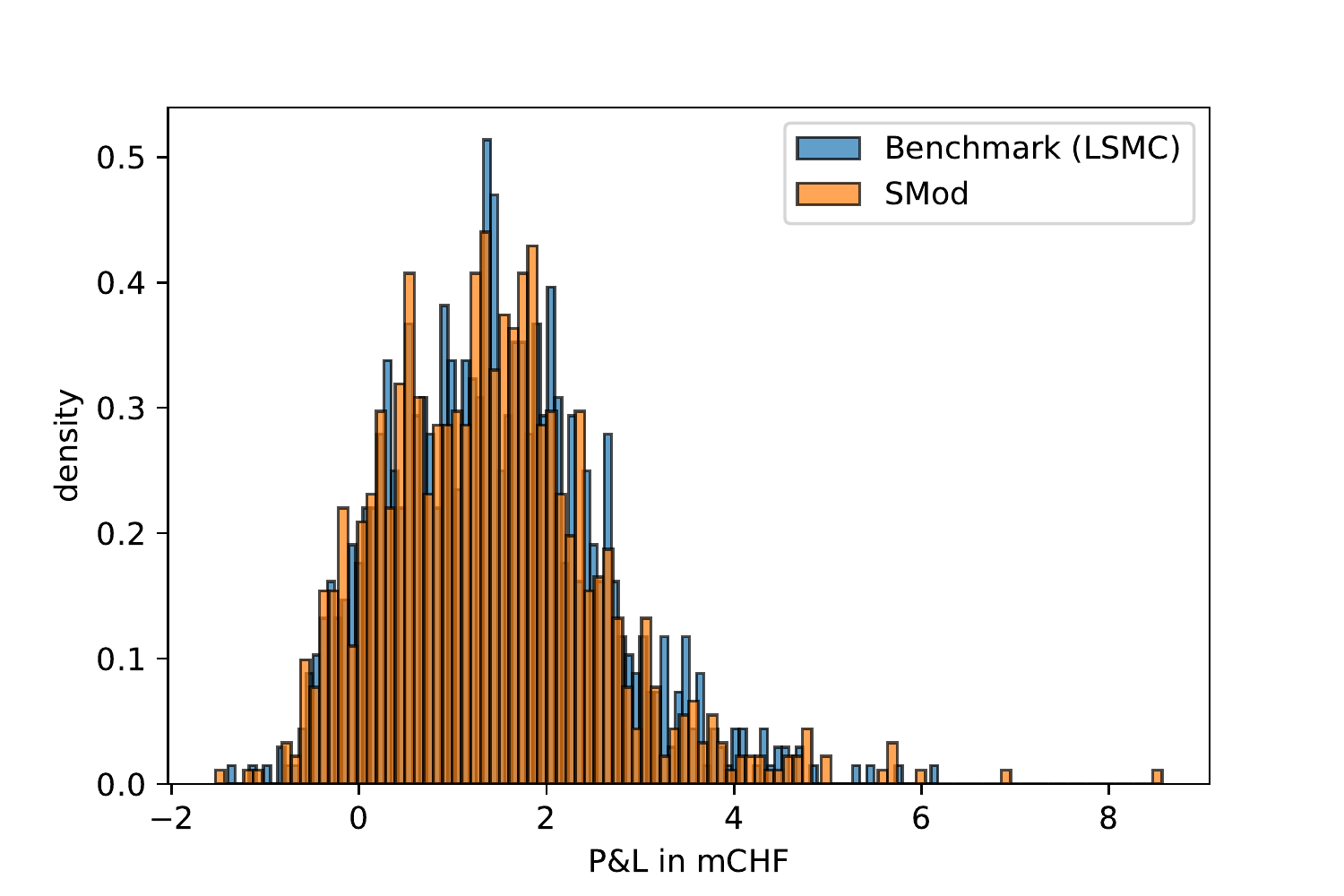}
	\hspace*{0.5em}
  	\includegraphics[trim={0cm 0cm 1.5cm 0cm},clip,width=0.46\textwidth]{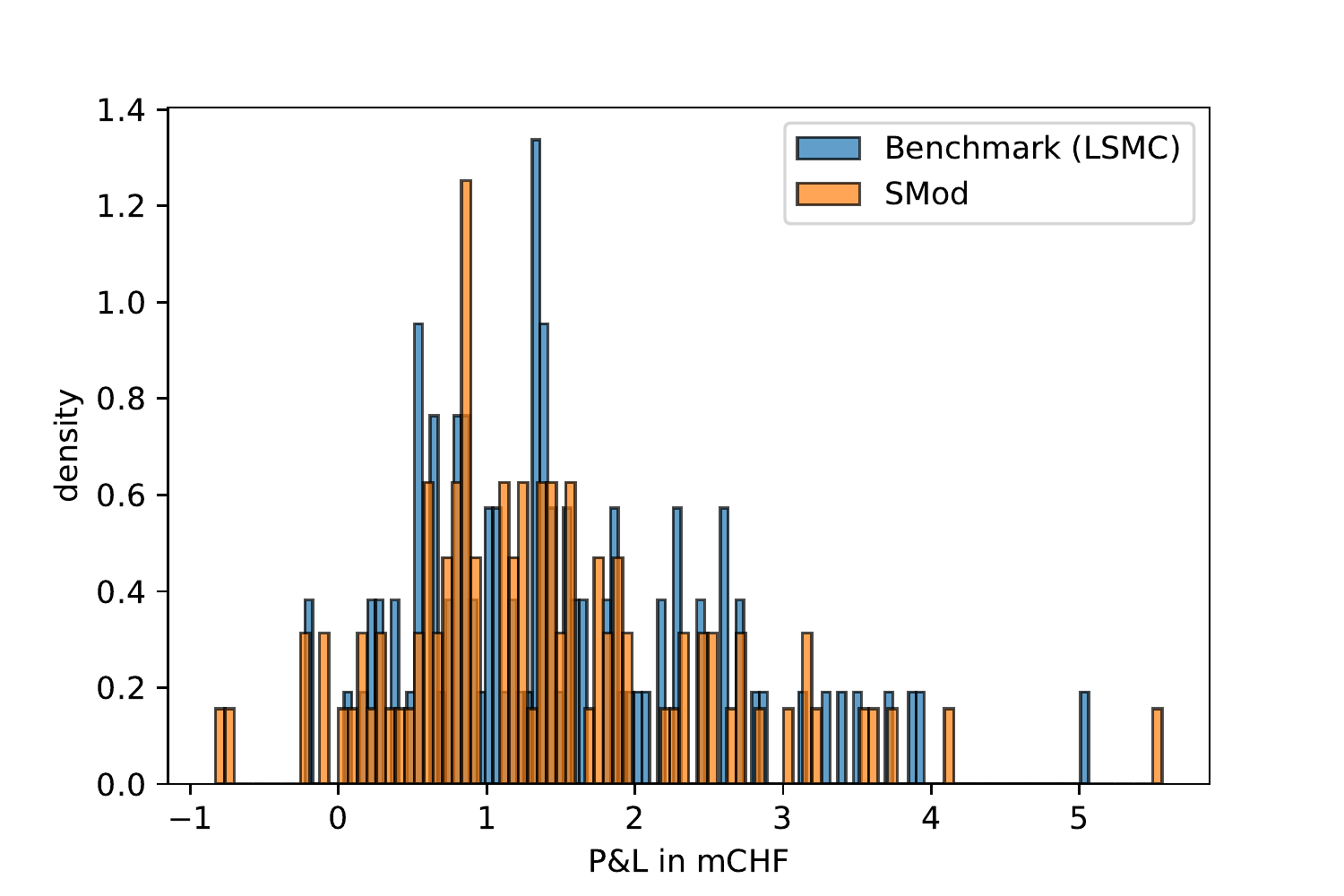}
    \caption{A line-up between the performance of the spot-only model (SMod) and that of the benchmark (LSMC). The upper left plot shows all the spot price scenarios. The the upper right plot shows the storage fill levels across all scenarios as inferred from the neural networks of SMod, which are similar to those of the benchmark actions. The scale is normalized by the storage capacity $c$. The optimal policy tends to inject gas until the storage capacity is reached, and withdraws it after a certain waiting period until the storage is empty again. This is in-line with the underlying seasonality pattern. The plots below compare the terminal P\&L between SMod and the benchmark in million CHF. The distribution on the left is based on the training set and that on the right based on the test set. Both in- and out-of-sample results are compellingly close to the benchmark.}
    \label{fig:comparison}
\end{figure}

\section{SFMod: intrinsic spot and forward trading}

In the following, we extend the previous model by trading additionally on the front month rolling forwards with delivery period of a whole month. A front month rolling forward curve contains at any point in time the first nearby monthly forward. We inherently assume that a monthly forward contract is only traded before its delivery period starts (and no longer during the delivery period), and that delivery obligations are valued using the spot prices whose delivery days lie within the delivery period. Note that restrict ourselves to those forwards that have delivery months within the time horizon of the storage problem. A visualization of the forward rolling mechanism is provided in \sref{Figure}{Fig_roll}.

\begin{figure}[H]
\centering
\begin{tikzpicture}[line cap=round, line join=round, x=4.5cm, y=4cm, set/.style={draw}] 
   \draw[color=darkgray,line width=0.4mm,->] (-0.05,0) -- (1.2,0) -- (1.22,-0.1) -- (1.24,0.1) -- (1.26,0) -- (2.15,0);
  \foreach \x in {0,...,6}
     	\draw (\x/5,1pt) -- (\x/5,-3pt);
  \draw (1.4,1pt) -- (1.4,-3pt);
  \draw (1.6,1pt) -- (1.6,-3pt);
  \draw (1.8,1pt) -- (1.8,-3pt);
 \draw (2,1pt) -- (2,-3pt);
  \draw[color=black] (0.05,-10pt) node[left] {$n_0$};
  \draw[color=black] (0.25,-10pt) node[left] {$n_1$};
  \draw[color=black] (0.45,-10pt) node[left] {$n_2$};
  \draw[color=black] (0.65,-10pt) node[left] {$n_3$};
  \draw[color=black] (0.85,-10pt) node[left] {$n_4$};
  \draw[color=black] (1.05,-10pt) node[left] {$n_5$};
  \draw[color=black] (1.7,-10pt) node[left] {$n_{10}$};
  \draw[color=black] (1.9,-10pt) node[left] {$n_{11}$};
  \draw[color=black] (2.1,-10pt) node[left] {$n_{12}$};
  \draw[-latex,thick] (0.5,0.4) node [right] { $\begin{array}{l} \triangleright\text{ Trade $h^S_k$ unit of $S_k$}\\ \triangleright \text{ Trade $h^1_k$ unit of $F(k,{\darkgreen n_1,n_2})$}\\ \triangleright \text{ No delivery obligation}\end{array}$} to [out=160,in=100,looseness=1.8] (0.1,0.02);
  \draw [draw = white!30!green,pattern = horizontal lines, pattern color = white!30!green] (0.2,0) rectangle (0.4,0.1); 
  \draw [pattern=north east lines] (0,0) rectangle (0.2,0.1);
 \end{tikzpicture}
 \begin{tikzpicture}[line cap=round, line join=round, x=4.5cm, y=4cm, set/.style={draw}]
   \draw[color=darkgray,line width=0.4mm,->] (-0.05,0) -- (1.2,0) -- (1.22,-0.1) -- (1.24,0.1) -- (1.26,0) -- (2.15,0);
  \foreach \x in {0,...,6}
     	\draw (\x/5,1pt) -- (\x/5,-3pt);
  \draw (1.4,1pt) -- (1.4,-3pt);
  \draw (1.6,1pt) -- (1.6,-3pt);
  \draw (1.8,1pt) -- (1.8,-3pt);
 \draw (2,1pt) -- (2,-3pt);
  \draw[color=black] (0.05,-10pt) node[left] {$n_0$};
  \draw[color=black] (0.25,-10pt) node[left] {$n_1$};
  \draw[color=black] (0.45,-10pt) node[left] {$n_2$};
  \draw[color=black] (0.65,-10pt) node[left] {$n_3$};
  \draw[color=black] (0.85,-10pt) node[left] {$n_4$};
  \draw[color=black] (1.05,-10pt) node[left] {$n_5$};
  \draw[color=black] (1.7,-10pt) node[left] {$n_{10}$};
  \draw[color=black] (1.9,-10pt) node[left] {$n_{11}$};
  \draw[color=black] (2.1,-10pt) node[left] {$n_{12}$};
  \draw[-latex,thick] (0.5,0.4) node [right] {$\begin{array}{l} \triangleright\text{ Trade $h^S_k$ unit of $S_k$}\\ \triangleright \text{ Trade $h^2_k$ unit of $F(k, {\darkgreen n_2, n_3})$}\\ \triangleright \text{ Daily delivery: $d^1$ unit}\end{array}$} to [out=160,in=100,looseness=1.8] (0.3,0.02);
  \draw [pattern=north east lines] (0.2,0) rectangle (0.4,0.1);
  \draw [draw = white!30!green,pattern = horizontal lines, pattern color = white!30!green] (0.4,0) rectangle (0.6,0.1);
\end{tikzpicture}
\begin{tikzpicture}[line cap=round, line join=round, x=4.5cm, y=4cm, set/.style={draw}]
   \draw[color=darkgray,line width=0.4mm,->] (-0.05,0) -- (1.2,0) -- (1.22,-0.1) -- (1.24,0.1) -- (1.26,0) -- (2.15,0);
  \foreach \x in {0,...,6}
     	\draw (\x/5,1pt) -- (\x/5,-3pt);
  \draw (1.4,1pt) -- (1.4,-3pt);
  \draw (1.6,1pt) -- (1.6,-3pt);
  \draw (1.8,1pt) -- (1.8,-3pt);
 \draw (2,1pt) -- (2,-3pt);
  \draw[color=black] (0.05,-10pt) node[left] {$n_0$};
  \draw[color=black] (0.25,-10pt) node[left] {$n_1$};
  \draw[color=black] (0.45,-10pt) node[left] {$n_2$};
  \draw[color=black] (0.65,-10pt) node[left] {$n_3$};
  \draw[color=black] (0.85,-10pt) node[left] {$n_4$};
  \draw[color=black] (1.05,-10pt) node[left] {$n_5$};
  \draw[color=black] (1.7,-10pt) node[left] {$n_{10}$};
  \draw[color=black] (1.9,-10pt) node[left] {$n_{11}$};
  \draw[color=black] (2.1,-10pt) node[left] {$n_{12}$};
  \draw[-latex,thick] (1.4,0.4) node [left] {$\begin{array}{l} \triangleright\text{ Trade $h^S_k$ unit of $S_k$}\\ \triangleright \text{ Trade $h^{11}_k$ unit of $F(k,{\darkgreen n_{11},n_{12}})$}\\ \triangleright \text{ Daily delivery: $d^{10}$ MWh.}\end{array}$} to [out=0,in=60,looseness=2.5] (1.7,0.02);
  \draw [pattern=north east lines] (1.6,0) rectangle (1.8,0.1);
  \draw [draw = white!30!green,pattern = horizontal lines, pattern color = white!30!green] (1.8,0) rectangle (2,0.1);
\end{tikzpicture}
\begin{tikzpicture}[line cap=round, line join=round, x=4.5cm, y=4cm, set/.style={draw}]
   \draw[color=darkgray,line width=0.4mm,->] (-0.05,0) -- (1.2,0) -- (1.22,-0.1) -- (1.24,0.1) -- (1.26,0) -- (2.15,0);
  \foreach \x in {0,...,6}
     	\draw (\x/5,1pt) -- (\x/5,-3pt);
  \draw (1.4,1pt) -- (1.4,-3pt);
  \draw (1.6,1pt) -- (1.6,-3pt);
  \draw (1.8,1pt) -- (1.8,-3pt);
 \draw (2,1pt) -- (2,-3pt);
  \draw[color=black] (0.05,-10pt) node[left] {$n_0$};
  \draw[color=black] (0.25,-10pt) node[left] {$n_1$};
  \draw[color=black] (0.45,-10pt) node[left] {$n_2$};
  \draw[color=black] (0.65,-10pt) node[left] {$n_3$};
  \draw[color=black] (0.85,-10pt) node[left] {$n_4$};
  \draw[color=black] (1.05,-10pt) node[left] {$n_5$};
  \draw[color=black] (1.7,-10pt) node[left] {$n_{10}$};
  \draw[color=black] (1.9,-10pt) node[left] {$n_{11}$};
  \draw[color=black] (2.1,-10pt) node[left] {$n_{12}$};
  \draw[-latex,thick] (1.6,0.4) node [left] {$\begin{array}{l} \triangleright \text{ Trade $h^S_k$ MWh of $S_k$}\\ \triangleright \text{ No action on forward market}\\ \triangleright \text{ Daily delivery: $d^{11}$ MWh.}\end{array}$} to [out=0,in=60,looseness=2.5] (1.9,0.02);
  \draw [pattern=north east lines] (1.8,0) rectangle (2.0,0.1);
  \draw [pattern=north east lines] (2.0,0) rectangle (1.8,0.1);
\end{tikzpicture}\caption{The mechanism of the rolling strategies in SFMod.}\label{Fig_roll}
\end{figure}
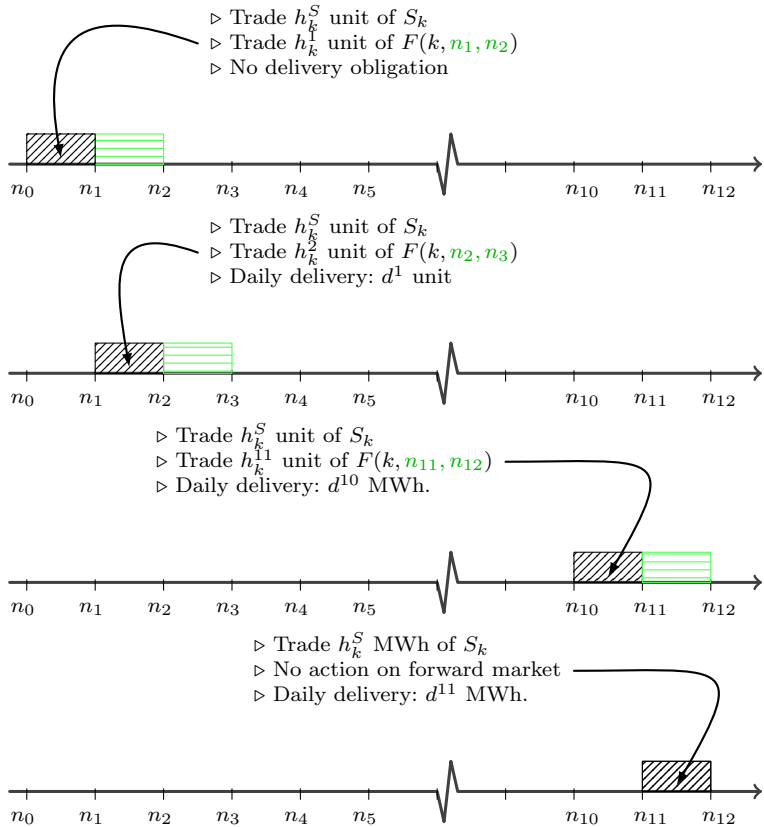

Let $0=n_0<n_1<...<n_J<K$ be the first days of the months $\mathcal{J}=\{0,1,...,J\}$ respectively. Let $h^j_k$ with $j\in\mathcal{J}$ denote the action on day $k$ on the forward $F(k,n_j, n_{j+1}-1)$, which has the delivery period $[n_j, n_{j+1}-1]$. $h^j_k>0$ refers to buying and $h^j_k<0$ refers to selling $F(k,n_j, n_{j+1}-1)$. The above assumption implies that $h^j_k=0$ for $k<n_{j-1}$ and for $k\geq n_j$; in particular $h^J_k=0$ for all $k\in\mathbb{T}$. Consistently to SMod, we aim to maximize 
\begin{align}\label{Eq_WTOyII}
\mathbb{E}_\mathbb{P} \big[U(W_{K-1})\big]
\end{align}
with the terminal P\&L
\begin{align*}
W_{K-1} := W^S_{K-1} + W^F_{K-1}.
\end{align*} 
$W^S_{K-1}$ denotes the terminal P\&L from the spot trading and is unchangedly given by \eqref{Eq_Wspot}. $h^S_k$ schedules the storage activity for the next day. Similarly, $W^F_{K-1}$ denotes the terminal P\&L from trading the monthly forward, and is defined as
\begin{align}\label{Eq_WF}
W^F_{K-1}= \sum_{j=1}^{J-1} ~\sum_{ k=n_{j-1} }^{n_j-1}\big(-h^j_k F(k, n_j, n_{j+1}-1)(n_{j+1}-n_{j})\big). 
\end{align} 
For a forward with the delivery period $[n_j, n_{j+1}-1]$, the daily delivery quantity $d^j$ is fixed on day $n_j -1$ for $j\geq 0$, and is given by
\begin{align}\label{eq:beginconstraintsmodel2}
    d^j:=\sum_{k=n_{j-1}}^{n_j-1} h^j_k,\qquad \text{for}~ j>0,
\end{align}
and $d^0:=0$. The storage level $H_n$ on day $n$ depends on both spot and monthly forward trading activities. For $n\in [n_{I-1},n_I)$, $H_n$ is given by
\begin{align*}
H_{n}:=\sum_{k=0}^{n-1}h^S_k + \sum_{j=1}^{I-2} \big( d^j (n_{J+1}-n_{J})\big)+d^{I-1} (n-n_{I-1}+1)
\end{align*}
with initially empty storage, i.e.\ $H_0:=0$. 
The optimization of \eqref{Eq_WTOyII} is subject to the constraints
\begin{align}
H_K & = 0,\\
0 \,\leq H_k \, \leq\, c, \qquad   &\text{and  }
\qquad\ell_k-d^j \, \leq \,h^S_k \, \leq\, u_k-d^j\label{Eq_contoyII}
\intertext{for $n_j\leq k\leq n_{j+1}$, $j\leq J-1$, and}
h^j_k &\leq \alpha\frac{c}{n_{j+1}-n_j} \label{Eq_fwdII}
\intertext{and for $\alpha\in[0,1]$. Alternatively, the daily constraints \eqref{Eq_contoyII} can be expressed as iterative daily bounds}
\widetilde{\ell}_k \, \leq \, h_k&+d^j\,  \leq \,\widetilde{u}_k,
\intertext{where}
\widetilde{\ell}_k:=\max\big\{\ell_k,-H_k\big\},\qquad &\text{and}\qquad\widetilde{u}_k:=\min\big\{u_k,c-H_k\big\}.\nonumber
\end{align}

\begin{rem}
In SFMod, the aggregated action on day $k$ is $(h^S_k+d^j)$ for all $n_j\leq k< n_{j+1}$. Hence, the action of pure spot trading is restricted by the daily delivery amount $d^j$, which results from $h^j_{\tilde{k}}$ with $n_{j-1}\leq\tilde{k}<n_j$. In other words, forward trading activities have a delayed effect on the spot trading, but spot trading does not affect forward trading. The delivery quantities of the upcoming days in the current month are fixed after the respective forward trading has already terminated. The delivery obligations of the due forwards restrict the spot trading activities of the current month, as the sum of daily delivery and the spot trading is bounded by the daily withdrawal and injection rates. The constraint \eqref{Eq_fwdII} ensures that the maximally traded amount can be stored in case of no spot trading. It can also be identified as liquidity constraint. Moreover, with the scaling factor $\alpha\in[0,1]$, one can bound the volume of forward trading and maintain an appropriate balance between spot and forward trading.
\end{rem}

\subsection{Training setup}
Similarly to Section \ref{ML_ToyI}, we approximate for each trading day $k$ actions $(h_k^S, h_k^j)$ by neural networks $g_k=(g_k^S, g_k^F)$ collected in $\widetilde{G}=\{g_0, ..., g_{K-1}\}$. Note that within this section, network strategies entail a two-dimensional output since in addition to actions in the spot market we also model strategies on the monthly forwards. For ease of notation, we abbreviate monthly forwards as $F_k=F(k, n_j, n_{j+1})$ for all $j\in\mathcal{J}$. The most important aspects of the training can be summarized as follows.
\begin{itemize}
\item Training data: time horizon of storage $\mathbb{T}$, $M$ trajectories of the spot $(S^i_k )_{k\in\mathbb{T};i=1,...,M}$, and of rolling month forward $(F^i_k)_{k\in\mathbb{T};i=1,...,M}$ respectively;
\item Training object: trading strategy network with two outputs for spot action and action in the rolling month forward consisting of $N\in\N$ ($N\le K$) distinct sub-networks, each of which has $L$ layers. The network's input is time as well as respective spot and storage fill level.
\item Training criterion: minimize an estimate of expected negative utility over batches $B\subset\{1,\ldots,M\}$ of training data, i.e.,
\[\min_{\widetilde{G}\in\mathcal{G}^i \forall i\in B}\frac{1}{|B|} \sum_{i\in B}U\big(W^{i,S}_{K-1} + W^{i,F}_{K-1} \big),\] 
where
\begin{align*}
W^{i,S}_{K-1} &:=\sum_{k=0}^{K-1}-g_k^S\big(k,H^{S^i}_k,S^i_k, F^i_k\big) S^i_k,\\
W^F_{K-1}&:= \sum_{j=1}^{J-1} ~\sum_{ k=n_{j-1} }^{n_j-1}\Big(-g^F_k\big(k,H^{S^i}_k,S^i_k, F^i_k\big) F^i_k(n_{j+1}-n_{j})\Big).
\end{align*}
For each scenario $i$, $\mathcal{G}^i$ contains those strategies $\widetilde{G}$ that fulfil all constraints \eqref{eq:beginconstraintsmodel2}--\eqref{Eq_fwdII}.
\end{itemize}

For numerical testing, $1\,000$ scenarios of spot as well as monthly forward price curves of 12 months were provided by Axpo~Solutions~AG. $1\,000$ rolling monthly forward curves were inferred, each of which contains only the first nearby contract on any trading day. As in Section~\ref{ML_ToyI}, the data set was split into $900$ training and $100$ test scenarios. Furthermore, we relied on the same network architecture as in SMod, i.e., 12 distinct neural networks representing 12 months of trading with $L=2$ and $d_1=16$. Note, however, that forward trading is discontinued in the last month, because the corresponding contract delivers beyond the trading horizon of the storage. Taking into account the revised constraints \eqref{eq:beginconstraintsmodel2}--\eqref{Eq_fwdII}, strategy networks were then trained for $1\,000$ epochs, using Adam stochastic gradient descent with a learning rate of $0.05$ and a batch size of $100$. \sref{Figure}{fig:storage} visualizes the resulting policy in terms of the storage level across the scenarios over time with respect to different choices of $\alpha$. \sref{Figure}{fig:alphacomparison} provides a visualization and detailed summary statistics for comparing this model with SMod and its benchmark. The comparison is based on the same setup for the spot strategy component and on the same training conditions with the exception of risk aversion rate\footnote{The risk aversion rate $r$ in the loss function can be used to shrink the variance of the P\&L. As the use of forwards in SFMod widens the P\&L variance significantly, a larger $r$ becomes necessary to control the distribution accordingly. We chose $r=3$ for SMod and $r=10$ for SFMod.}. Compared with the spot-only model, SFMod entails not only a significantly higher P\&L on average, but also a higher volatility. Moreover, the P\&L originates mostly from forward trading activities, and thus, it is highly sensitive with respect to the choice of $\alpha$ in \eqref{Eq_fwdII}. Hence, a suitable choice of $\alpha$ is inevitable.

\begin{figure}[H]
    \centering
	\includegraphics[trim={0cm 0cm 1.5cm 0cm},clip,width=0.46\textwidth]{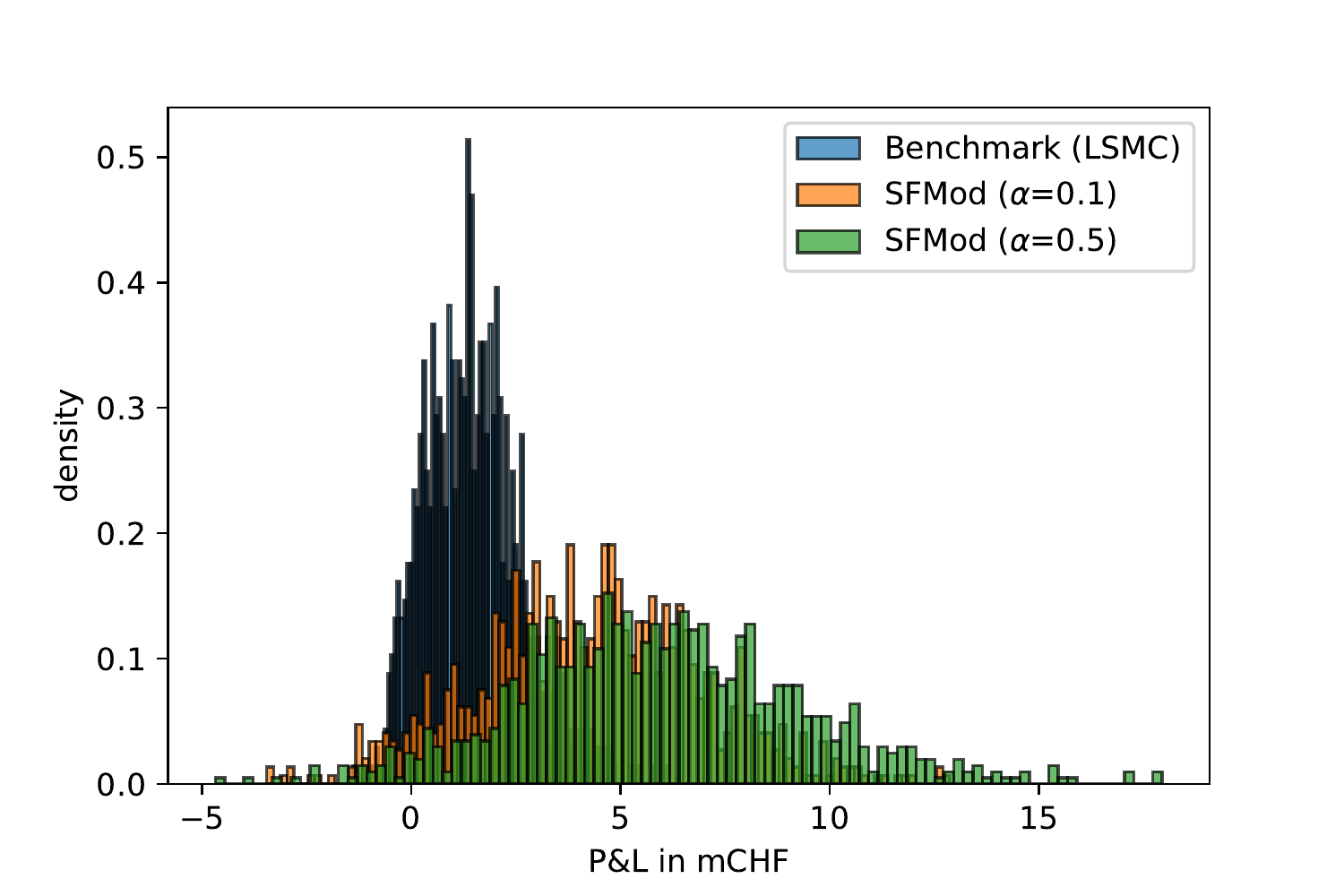}
	\hspace*{0.5em}
	\includegraphics[trim={0cm 0cm 1.5cm 0cm},clip,width=0.46\textwidth]{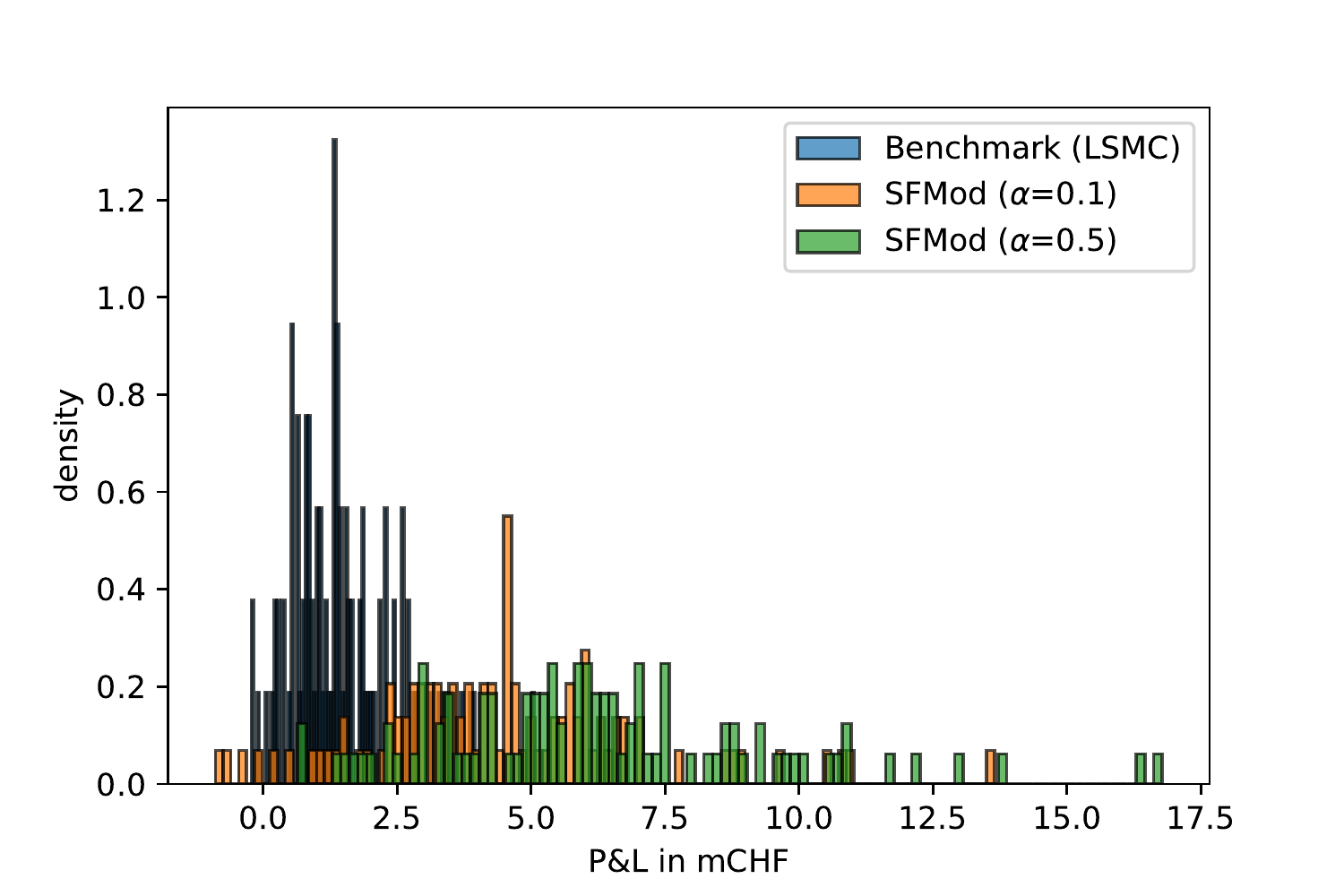}
    \caption{A comparison between the performance of SFMod with different choices for $\alpha$ and that of the benchmark (LSMC). The plots compare the terminal P\&L between SFMod ($\alpha= 0.1$ and $\alpha=0.5$) and the benchmark in million CHF. The left chart exhibits the P\&L on the training set and the right chart that on the test set. In both plots, the P\&L distributions of SFMod are conclusively more favorable than that of the benchmark. The optimal policy features a consistent seasonality pattern as that in SMod.}
    \label{fig:storage}
\end{figure}

\sref{Figure}{fig:alphacomparison} substantiates that SFMod, which allows trading activities on forwards, is clearly the most favorable choice in terms of maximizing the expected utility of terminal wealth. In comparison to SMod, it is slightly more involved from the technical setup, but in terms of computational time and effort, it is still well manageable on a standard 8-core notebook. Regarding SFMod, the higher first moment of the P\&L distribution across all scenarios comes with a higher standard deviation. Furthermore, the first moment is sensitive to the limitation on forward market activities, expressed by the control variable $\alpha$. One direction of future work might be to generate superior P\&L distributions with less risk. Another possible direction of future work might increase the model-theoretic complexity with more forward curves. It needs to be noted that the performance of SFMod is not adversely affected if we further extend the scope of forward trading activities or incorporate more realistic model features such as, for instance, $H^{S^i}_k$-dependent transaction cost.

\begin{figure}[H]
    \centering
\includegraphics[trim={0cm 0cm 0cm 0cm},clip,width=0.8\textwidth]{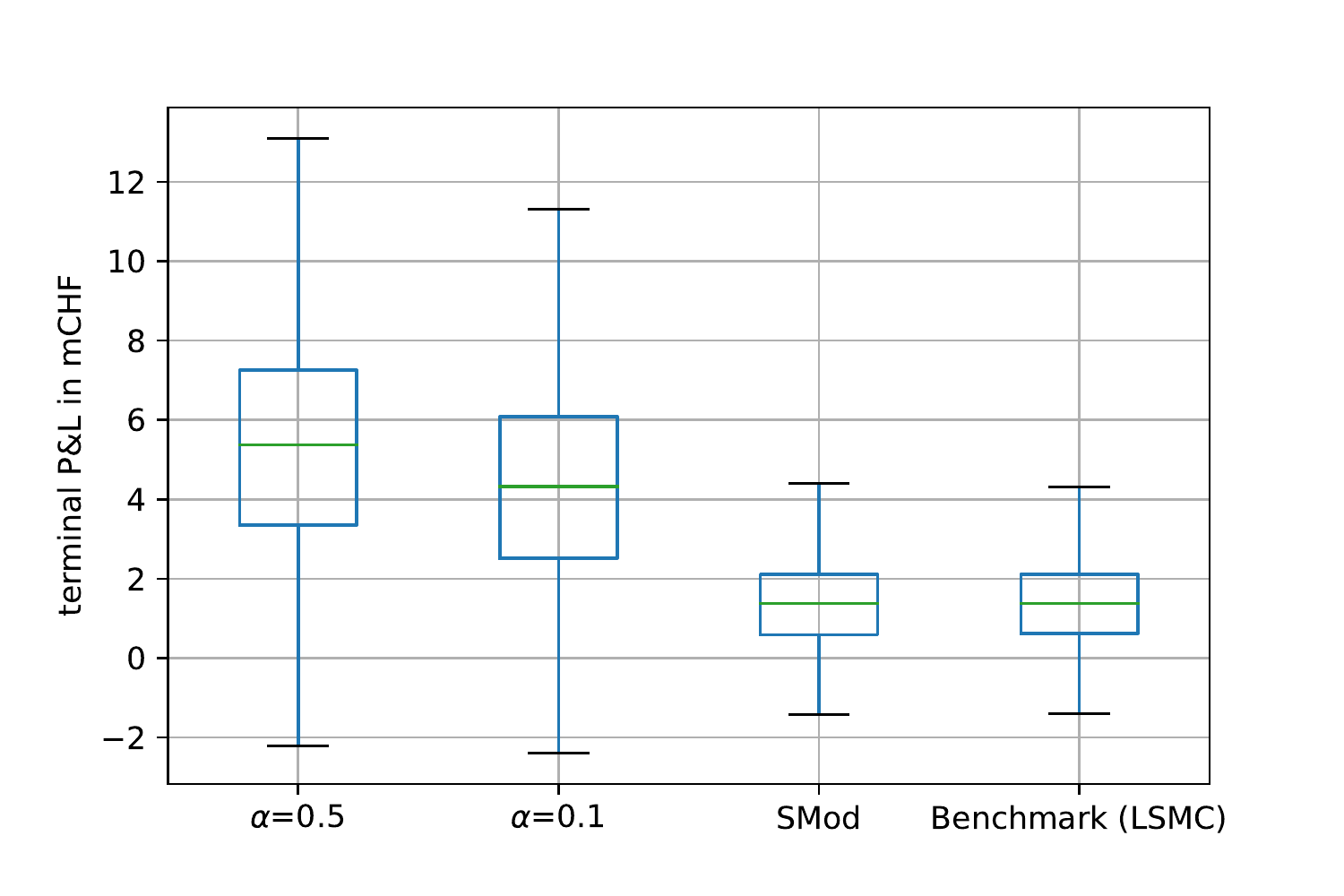}
\resizebox{0.85\columnwidth}{!}{%
 \begin{tabular}{c| c|c|c |c|c|c |c }
  \multirow{2}{*}{terminal P\&L in mCHF} & \multicolumn{3}{c|}{training set} & \multicolumn{3}{c|}{test set} & benchmark\\ \cline{2-8}
 & $\alpha=0.5$ & $\alpha=0.1$ & SMod & $\alpha=0.5$ & $\alpha=0.1$ & SMod & spot only\\ \hline \hline 
 average & 5,429,539 &	4,284,428 &	1,440,326 &	5,655,886 &	4,425,810 & 1,436,510  & 1,448,475 \\ \hline
 median & 5,362,543	& 4,341,783	& 1,391,877	& 5,576,047 & 	4,258,077 & 1,284,445 & 1,382,435 \\ \hline 
 std  & 3,111,824 &	2,708,985 &	1,132,284 &	2,839,238 &	2,526,534 &  1,019,118 & 1,109,619 \\ \hline
 \end{tabular}}
  \caption{The boxplot and the table provide a line-up of the considered models. The first moments and the volatility of the P\&Ls distribution are largest for SFMod. Moreover, they depend on the choice of $\alpha$. Please note that the LSMC-approach may not serve as a valid and competitive benchmark, as it does not allow for forward trading activities.}
    \label{fig:alphacomparison}
\end{figure}

\section{Conclusion}

We proposed a flexible and powerful framework that is capable of dealing with the intricacy of optimizing underground gas storage facilities in the presence of forward markets. Traditional techniques such as, for instance, least-squares Monte~Carlo (LSMC) or dynamic programming are subject to a so-called curse-of-dimensionality, whereas the proposed deep learning technique is almost not affected by the dimensionality. Moreover, our experimental results show that the proposed deep hedging approach performs as good or better than the most-established state-of-the-art LSMC benchmark. These advances pave the way for unprecedented storage and production plans of energy.


%
%


\bibliographystyle{plainnat}
\bibliography{Bibliographie}

\end{document}